\documentclass[lettersize,journal]{IEEEtran} 
\usepackage{cite}
\usepackage{mathrsfs}
\usepackage{amsthm}

\usepackage[cmex10]{amsmath}
\usepackage{mathtools}
 \usepackage{siunitx}
\usepackage{array}
\usepackage{eqparbox}
\usepackage{bm}

\usepackage[table,dvipsnames]{xcolor}
\usepackage{multicol,booktabs,tabularx}

\usepackage{cases}
\usepackage{algorithm}
\usepackage{paralist}
\usepackage{algpseudocode}

\usepackage{graphicx}
\usepackage{subfigure}
\usepackage{epstopdf}
\usepackage{epsfig}
\usepackage{amssymb}
\usepackage{array}
\usepackage{multirow}

\begin{document}

\title{Sidelink Positioning: Standardization Advancements, Challenges and Opportunities}

\author{Yuan Gao, \textit{Member, IEEE}, Guangjin Pan, Zhiyong Zhong, Zhengyu Jin, Yichen Hu, Yifei Jin, \\Shugong Xu, \textit{Fellow, IEEE}
\thanks{This work was supported by Shanghai Natural Science Foundation under Grant 25ZR1402148. (Shugong Xu is the corresponding author)} 
\thanks{Yuan Gao and Zhiyong Zhong are with the School of Communication and Information Engineering, Shanghai University, China, email: gaoyuansie@shu.edu.cn and zzy20010120@shu.edu.cn.}
\thanks{Guangjin Pan is with the Department of Electrical Engineering, Chalmers University of Technology, Gothenburg, Sweden, email: guangjin.pan@chalmers.se.}
\thanks{Yichen Hu is with the Department of Information Engineering, The Chinese University of Hong Kong, China, email: yichen@link.cuhk.edu.hk.}
\thanks{Zhengyu Jin is with the Department of Electrical and Computer Engineering, University of Washington, US, email:zjin25@uw.edu.}
\thanks{Yifei Jin is with the KTH Royal Institute of Technology \& Ericsson AB Division of Theoretical Computer Science \& Ericsson Research, Stockholm, Sweden, email: yifeij@kth.se.}
\thanks{Shugong Xu is with Xi’an Jiaotong-Liverpool University, Suzhou, China, email: shugong.xu@xjtlu.edu.cn.}
}

\maketitle

\begin{abstract}
With the integration of cellular networks in vertical industries that demand precise location information, such as vehicle-to-everything (V2X), public safety, and Industrial Internet of Things (IIoT), positioning has become an imperative component for future wireless networks. By exploiting a wider spectrum, multiple antennas and flexible architectures, cellular positioning achieves ever-increasing positioning accuracy. Still, it faces fundamental performance degradation when the distance between user equipment (UE) and the base station (BS) is large or in non-line-of-sight (NLoS) scenarios. To this end, the 3rd generation partnership project (3GPP) Rel-18 proposes to standardize sidelink (SL) positioning, which provides unique opportunities to extend the positioning coverage via direct positioning signaling between UEs. Despite the standardization advancements, the capability of SL positioning is controversial, especially how much spectrum is required to achieve the positioning accuracy defined in 3GPP. To this end, this article summarizes the latest standardization advancements of 3GPP on SL positioning comprehensively, covering \begin{inparaenum}[\itshape a\upshape)]
\item network architecture;
\item use cases;
\item positioning types; and
\item performance requirements.
\end{inparaenum} The capability of SL positioning using various positioning methods under different imperfect factors is evaluated and discussed in-depth. Finally, according to the evolution of SL in 3GPP Rel-19, we discuss the possible research directions and challenges of SL positioning.
\end{abstract}
\section{Introduction}
Positioning capabilities in mobile networks have advanced over many years, delivering accuracy that ranges from tens of meters for urgent calls to mere fractions of a meter indoors\cite{pan2025ai,gao2024performance}. These improvements address the rising need for exact location data across multiple fields. Sectors like vehicle communications, emergency response, industrial connectivity, and everyday uses increasingly depend on network-based positioning\cite{22_261}. These applications benefit from network upgrades, such as expanded frequency bands, multiple antennas, and adaptable designs\cite{xu2025enhanced,gao2025stochastic}.

Yet, obstacles including equipment flaws, signal reflections, and obstructed direct paths\cite{xu2025enhanced} prevent these systems from meeting strict standards set by the 3rd Generation Partnership Project (3GPP) for precision, reliability, and speed\cite{38_857}. One approach involves adding more network points to shorten distances between devices and stations, thus raising the chances of line-of-sight (LoS). Still, such expansion substantially raises deployment costs, motivating alternative approaches to improve positioning performance.

To fulfill these stringent positioning requirements of 3GPP, direct device-to-device positioning through sidelink (SL) connections has emerged and is under standardization to improve overall network performance\cite{jornod2025positioning}. This allows nearby devices to exchange location signals directly, cutting down separation and enhancing direct visibility\cite{ge2024v2x}. It also enables shared positioning at lower expense than dense station setups, assuming enough devices are present. Moreover, sidelink can provide location services where device-to-station links are weak or absent, like in partial coverage and out-of-coverage areas\cite{38_859}.

To meet the stringent positioning requirements defined by 3GPP, sidelink (SL) positioning via the PC5 link has been proposed and is currently being standardized by 3GPP to enhance cellular positioning performance. By enabling direct communication of position reference signals between UEs, the distance between anchors and UEs can be significantly reduced, thereby increasing the probability of LoS paths\cite{ge2024v2x}. Furthermore, cooperative positioning can be achieved more cost-effectively compared to densely deployed BSs, provided there are a sufficient number of UEs. Additionally, SL can offer positioning services in areas where communication between UEs and BSs is unavailable, such as in partial coverage and out-of-coverage areas\cite{38_859}. 

Even with these developments, a comprehensive overview of recent standardization progress and requirements through Release 19 remains missing. The technical aspects of SL positioning are fragmented in various 3GPP technical reports, hindering a holistic understanding of the subject. Earlier works\cite{le20235g,ganesan20235g} overlook the structure of SL in Release 18 and fail to examine Release 19 needs deeply. Further, vendors disagree on what SL can realistically achieve. One analysis\footnote{Huawei, “Finalizing SL positioning evaluation,” 3GPP Work Item Description, Tech. Rep. R1-2210900 (2022-11), 2022.\label{FN:HW}} finds that in highway vehicle scenarios, SL alone with 100 MHz bandwidth misses the 1.5-meter relative precision at 90\% reliability, while another insists it succeeds under the same setup\footnote{ZTE, “Summary 1 for SL positioning evaluation,” 3GPP Work Item Description, Tech. Rep. R1-2211506 (2022-11), 2022.\label{FN:ZT}}. Urban layouts show similar disputes over hitting 1.5 meters with 100 MHz sidelink\textsuperscript{\ref{FN:HW},\ref{FN:ZT}}. Moreover, factors such as clock drift and asynchronization have not been considered in these simulations, further complicating the assessment of SL positioning capabilities. 

Accordingly, this article compiles key elements from the latest technical reports and standards, offering a unified view of advancements in Releases 18 and 19. It covers system structures, resource distribution, positioning approaches, and positioning reference signals (PRSs) in Release 18. We contrast requirements between Releases 18 and 19, pointing out hurdles for future networks. By reviewing the technical reports of 3GPP, we provided lessons regarding the capabilities of SL positioning and the contracting performance evaluation among vendors. Our examination delves into the root causes of these differences. By bridging standardization efforts with practical deployment considerations, this paper aims to offer insights and research directions specifically to SL positioning for 5G and beyond.

\begin{figure*}
  \centering
  \includegraphics[width=1\textwidth]{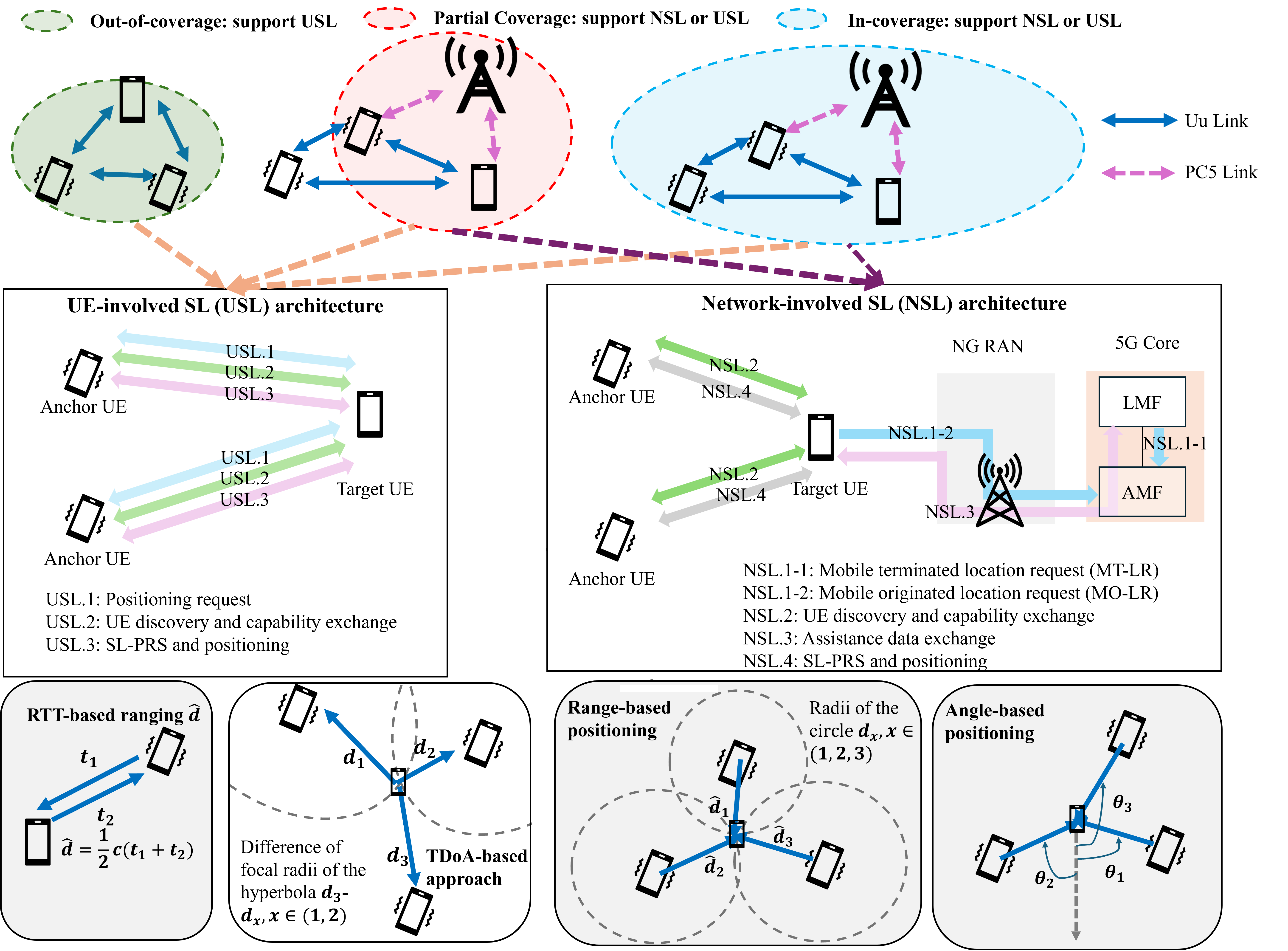}
  \caption{The scenarios of SL positioning, semantic diagram of network-involved SL (NSL) and UE-only SL (USL) architecture and the SL positioning methods. }
  \label{scenario_procedure}
\end{figure*}

%In the Section II of this paper, we first summarize the latest standardization advancements of 3GPP on SL positioning comprehensively, covering SL positioning architectures, resource allocation, positioning types, and positioning reference signals (PRSs) in 3GPP Rel-18. We also compare the positioning requirements between 3GPP Rel-18 and Rel-19. In Section III, we examine the simulation settings used by different operators and manufacturers that lead to conflicting results and offer recommendations for SL positioning simulations. In Section IV, we analyze the impact of various influential factors on SL positioning, including radio resources, the number of anchors, and imperfect factors. In Section V, following the trajectory of SL standardization in 3GPP Rel-19, we identify and discuss future research directions and opportunities for SL positioning. The paper concludes with a summary of our findings in Section VI.
\section{A Holistic Picture of SL Positioning in 3GPP}

\subsection{A Brief Overview of Cellular Positioning}
The evolution of positioning technology in mobile communication systems has progressed from coarse, cell-based methods to highly accurate, multi-faceted approaches. This advancement has broadened the scope of applications and significantly enhanced the capabilities of mobile positioning systems\cite{ganesan20235g}.

Initially, mobile communication relied on Cell-ID for positioning, using the unique identifier of the connected cell tower to estimate a device's location, which only provides rough location within several tens of meters. Over time, more advanced techniques, such as range-based and angle-based estimation, have been developed to improve positioning performance. In 3GPP Rel. 18, the following positioning enhancements are proposed: (1) flexible PRS bandwidth allocation up to 400 MHz to improve temporal resolution for time-of-arrival (ToA) and round-trip time (RTT), measurements; (2) hybrid positioning schemes combining RTT, time
difference of arrival (TDoA), and angle-of-arrival (AoA) with fusion filters to exploit measurement diversity; (3) support for both in-coverage and out-of-coverage operation with infrastructure-assisted anchors (e.g., RSUs, BSs) to improve synchronization and anchor geometry. 

However, practical cellular networks face challenges to achieve ubiquitous high-precision positioning: \begin{inparaenum}[\itshape a\upshape)]
\item typical scenarios that require positioning services, such as the indoor factories (InF), have lots of NLoS cases due to the complex physical environment full of scatters and obstacles. Cellular positioning accuracy degrades significantly in these NLoS cases;
\item cellular positioning is not available for UEs out of the coverage of cellular network, especially for public safety, such as the cases of fire emergency and after-earthquake rescue, where cellular services are out of order.
\end{inparaenum}

\subsection{SL Positioning Architectures}
To address these issues, SL positioning, which is achieved by transmitting and receiving PRSs between UEs, has been standardized by 3GPP in Rel-18. This section provides an overview of SL positioning architectures, including entity types, frameworks, and resource allocation schemes. We further explore both network-involved and UE-only scenarios, highlighting their characteristics and applications.

\subsubsection{Entity Types in SL Positioning}
In the SL positioning system, there are several types of entities: UE, access and mobility management function (AMF), location management function (LMF), and gateway mobile location center (GMLC). UEs are the devices that can be targets for localization, reference anchors for positioning, or servers for SL operations. AMF manages location capability information, such as the number of antennas and computation power of UEs, and forwards location requests to LMF. LMF oversees SL positioning services, processes requests, and performs calculations. GMLC interacts with user data management (UDM) to handle privacy settings and determine the appropriate AMF for location services.

\subsubsection{SL positioning scenarios}
SL positioning can be applied in various network coverage scenarios as in Fig. \ref{scenario_procedure}, each with its unique characteristics. In the in-coverage scenarios, all the UEs are within the cellular BS coverage area. In the partial coverage scenarios, one UE is within the BS coverage, while the other is outside. In the out-of-coverage scenarios, all the UEs are outside the BS coverage area. Depending on whether the communication between UE and BS is available, 3GPP categorizes SL positioning into \textit{network-involved SL (NSL) positioning} and \textit{UE-only SL (USL) positioning}. Illustrations of in-coverage, partial-coverage and out-of-coverage scenarios can be found in Fig. \ref{scenario_procedure}.
%The former is applicable for in-coverage or partial-coverage scenarios, requiring access to the 5G Core (5GC). The latter is used in out-of-coverage scenarios or when network support is unavailable. 
\begin{itemize}
\item In NSL positioning, the 5GC assists in service request processing and result calculations in the following stages. 
\begin{inparaenum}
\item \textbf{NSL.1:} SL positioning can be initiated either by the AMF (NSL.1-1) or UE (NSL.1-2), denoted as mobile terminated location request (MT-LR) or mobile originated location request (MO-LR), respectively. The latter involves the GMLC for privacy protection. 
\item \textbf{NSL.2:} The target UE selects anchor UEs participating in the positioning and establishes a secure link with them. Further, UEs exchange their positioning capabilities (information about supported positioning methods, parameters, etc.) through the SL positioning protocol.
\item \textbf{NSL.3:} The target UE requests assistance data and the method of transmitting location results from AMF. AMF selects LMF to send assistance information request, and LMF sends assistance data to the UE for positioning measurement.
\item \textbf{NSL.4:} The target UE performs the positioning task using collected measurement results based on the PRSs.    
\end{inparaenum}
\item USL positioning is employed when: \begin{inparaenum}[\itshape a\upshape)]
\item both target and anchor UEs are disconnected from the 5GC;
\item the 5GC does not support network-based SL positioning;
\item network congestion prevents SL positioning requests.
\end{inparaenum} 
In this architecture, an SL positioning server (often the target UE) manages the positioning process. The procedure is similar to network-involved SL positioning but relies solely on SL communication for device discovery, capability exchange, and result computation. USL positioning simplifies communication paths but may result in lower positioning accuracy compared to network-involved methods. For absolute positioning, reliable anchor UE coordinates must be provided through the network.
\end{itemize}

\begin{table*}[]
    \caption{A summary of the major influential factors, latency and required bandwidth for each positioning techniques}
    \label{pos_methods}
\begin{tabular}{|l|l|l|l|l|l|}
\hline
& {Positioning method} & {Positioning type} & {Major influential factors} & {Latency} & Bandwidth Required\\ \hline
\multirow{3}{*}{\begin{tabular}[c]{@{}l@{}}Time-based \\ approach\end{tabular}} & Single-sided RTT & \multirow{2}{*}{\begin{tabular}[c]{@{}l@{}}Ranging\end{tabular}} & Clock drift & \multicolumn{1}{l|}{Medium} &  Medium\\ \cline{2-2} \cline{4-6} 
 & Double-sided RTT &  & Clock drift & \multicolumn{1}{l|}{High} &Medium\\ \cline{2-6} 
 & TDOA & Primarily absolute positioning & Synchronization between UEs & \multicolumn{1}{l|}{Low} &  High\\ \hline
\multirow{2}{*}{\begin{tabular}[c]{@{}l@{}}Angle-based \\ approach\end{tabular}} & AOA & \multirow{2}{*}{Absolute/relative positioning} & \multirow{2}{*}{\begin{tabular}[c]{@{}l@{}}Number of antennas, arrangement\end{tabular}} & \multicolumn{1}{l|}{Low} &  Low\\ \cline{2-2} \cline{5-6} 
 & AOD &  &  & \multicolumn{1}{l|}{Low} & Low \\ \hline
\end{tabular}

\end{table*}

\subsection{Positioning Performance Requirements}

\begin{figure*}
  \centering
  \includegraphics[width=1\textwidth]{requirements.png}
  \caption{Positioning performance evolution from 3GPP Rel-18 to Rel-19\cite{22_261}. PSL is short for positioning service level. $\square$ 1-6 and $\square$ 7 apply to absolute and relative positioning, respectively. $\bigcirc$ 1-4 apply to both absolute and relative positioning. Maximum mobility supported for each positioning case: a) MI$xx$: indoor $xx$ km/h; b) MD$xx$: dense urban $xx$ km/h; c) MO$xx$: outdoor $xx$ km/h; d) MT$xx$: train $xx$ km/h.}
  \label{pos_requiurements}
\end{figure*}

The positioning performance requirements from 3GPP Rel-18 to Rel-19 are summarized in Fig. \ref{pos_requiurements}, and the most significant evolutions are the expansion of use cases, improved positioning performance, and the introduction of latency requirement. 
\subsubsection{Use Cases Expansion}
In 3GPP Rel-18, four use cases are introduced, namely V2X, public safety, IIoT, and commercial applications, where horizontal, vertical positioning accuracy and angle accuracy are defined with mobility up to \SI{30}{\kilo\metre/\hour}\cite{38_859}. In 3GPP Rel-19\cite{22_261}, the use cases are further generalized into 7 positioning service levels (PSLs), which can be classified as the indoor and outdoor scenarios. Public safety, IIoT and indoor commercial applications are typical indoor scenarios, of which mobility requirements are consistent with those defined in 3GPP Rel-18\cite{38_859}. At the same time, the outdoor scenarios (both rural and urban) support the positioning of vehicles and trains with speeds up to \SI{250}{\kilo\metre/\hour} and \SI{500}{\kilo\metre/\hour}, respectively. The 5G enhanced positioning service area further supports positioning in dense urban (up to \SI{60}{\kilo\metre/\hour}), tunnels for both vehicles and railways.

\subsubsection{Positioning Improvement and Latency}
Apart from horizontal and vertical positioning accuracy, positioning service availability and positioning service latency are introduced in 3GPP Rel-19 to support mission-critical services\cite{22_261}. A total of 7 positioning requirements are defined with 6 levels and 1 level of relative positioning accuracy requirement. In the absolute positioning accuracy requirement sets, horizontal accuracy and vertical accuracy range from \SI{10}{\metre} to \SI{0.3}{\metre}, and \SI{2}{\metre} to \SI{3}{\metre}, respectively. Positioning service availability ranges from $95$ \% to $99.9$ \%. Latency at the one-second level is required to support outdoor scenarios, i.e., PSLs 1, 2, 3 and 5. It should be noted that positioning with $ms$-level latency, i.e., PSLs 4 and 6, is only required in the indoor scenarios with 5G enhanced positioning service, such as the robot cooperation and collision avoidance of mobile robot (AMR) and in-factory scenarios. The relative positioning accuracy requirement, i.e., PSL 7, is required for the positioning of two UEs within $10$ m.

\section{SL positioning methods}

In 3GPP Rel-18, SL positioning methods are categorized into time-based and angle-based approaches, enabling absolute or relative positioning with high accuracy. Time-based methods are essential for supporting RTT and TDoA estimations. In contrast, angle-based methods are required to support at least AoA and AoD estimations. A systematic comparison of these positioning methods is summarized in Table \ref{pos_methods}.
\subsection{RTT Method}
As illustrated in Fig. \ref{scenario_procedure}, in RTT methods, two UEs transmit and receive SL PRSs to estimate the distance between them by calculating the time difference between signal transmission and reception. In the single-sided RTT approach, a simplified version, each UE transmits an SL-PRS signal only once, and the range is estimated by measuring the time of flight of the SL-PRS signal. However, clock drift within each UE can introduce errors in the positioning measurement. To mitigate the impact of clock drift at the expense of positioning latency, the double-sided RTT method is employed, where one UE transmits twice while the other UE transmits once.

\begin{figure*}
  \centering
  \includegraphics[width=1\textwidth]{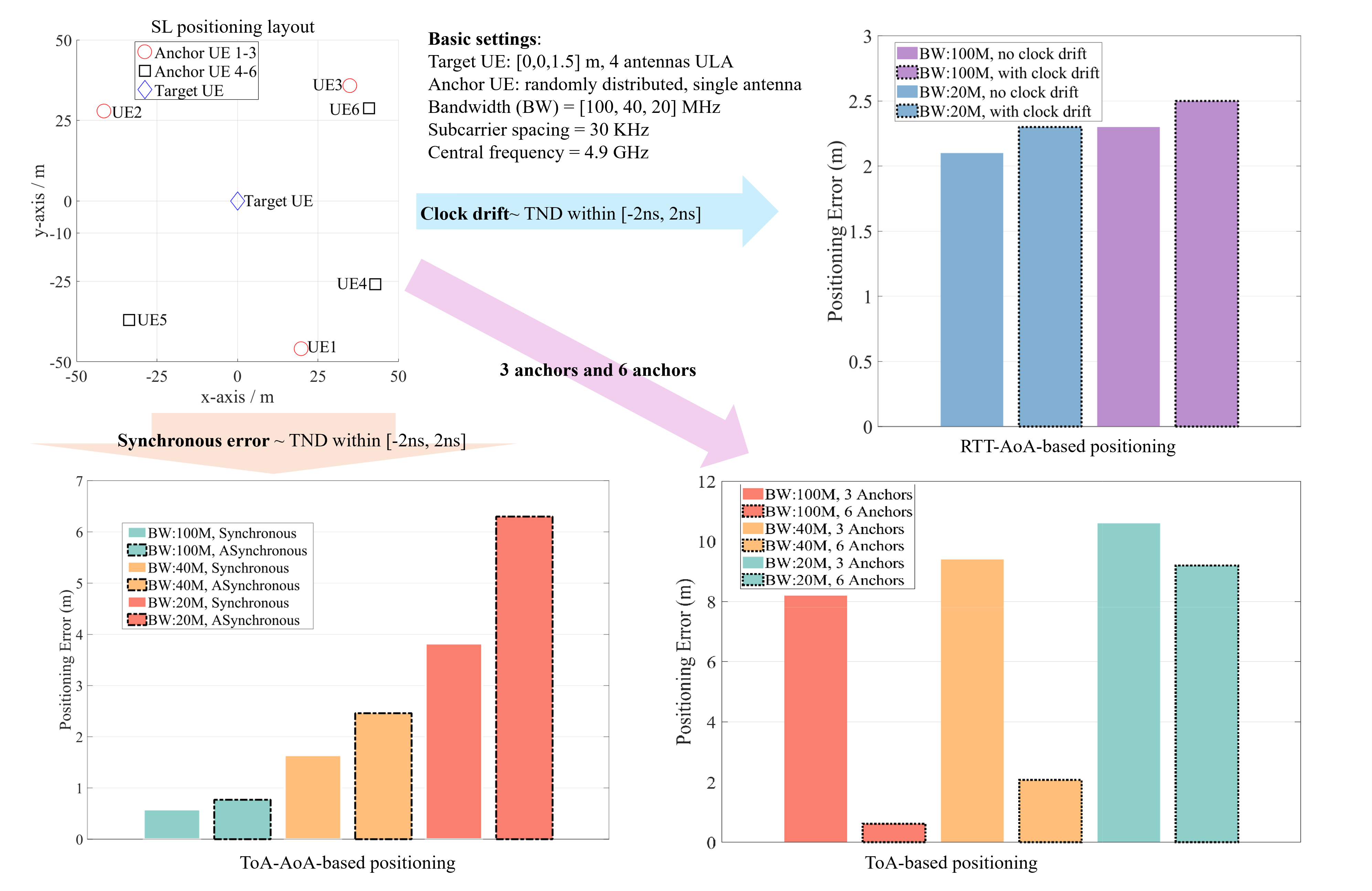}
  \caption{Illustrations of SL positioning layout, simulation settings and simulation results in terms of 90\%-percentile positioning error. TND is short for truncated normal distribution.}
  \label{A_AS_90_ToA_AoA}
\end{figure*}

\subsection{TDOA-Based Approach}
As shown in Fig. \ref{scenario_procedure}, in TDoA-based SL positioning, a UE determines the target’s location by measuring the arrival times of signals from multiple nodes and calculating the differences between these times. Implementing SL-TDoA requires at least four UEs: one UE measures the ToA of signals from the SL-PRS, while three others transmit these signals. The estimated position of the target UE lies at the intersection of hyperbolas defined by the time differences, with each hyperbola characterized by a difference in focal distances of $d_3 - d_x, x\in[1,2]$, where $d_3$ is the maximum distance from the target UE to an anchor UE. Although achieving precise synchronization among UEs can be challenging, SL-TDoA remains a practical choice because it does not require two-way signaling between transmitters and receivers. This approach is primarily suited for absolute positioning, especially when the locations of anchor UEs are known. In such cases, the synchronization process is simplified, enhancing the method's viability.

\subsection{Angle-Based Approach}

SL-AoA and SL-AoD rely on angle measurements and are immune to time synchronization errors. SL-AoA includes measurements for both azimuth and zenith angles of arrival, with UEs equipped with multiple antenna elements to facilitate these measurements. The accuracy of the angle of arrival measurement depends on factors such as the number of antennas at the device, their spatial arrangement, and the relative distances between these antennas. SL-AoA is used for both absolute and relative positioning. In contrast, SL-AoD relies on measurements from SL-PRS at the transmitting UE, making it less prioritized than the other three methods due to the limited number of antennas at the UE.

\section{Lessons learned from 3GPP technical reports}

In this section, we examine the performance of SL positioning as detailed in 3GPP technical reports. Our analysis reveals that reaching 0.5-meter accuracy with 90\% availability, in the absence of base stations or roadside units, poses significant difficulties. We also identify the main factors contributing to these challenges. Given contrasting performance evaluation among vendors, we analyze the reason to help foster consensus in the field.

\subsection{Key finding 1: Achieving 0.5-meter positioning accuracy of 90\% availability without BS or RSU is challenging} 

A critical benchmark for V2X communications in Rel. 18 is 0.5-meter positioning accuracy at 90\% availability, and a number of vendors have simulated scenarios to evaluate this. The technical reports from 3GPP suggest that attaining this level without BSs or RSUs is difficult. When synchronization errors are present, methods such as RTT-based\textsuperscript{\ref{FN:HW}}, combined RTT and AoA-based\footnote{Samsung, “Discussion on evaluation for SL positioning,” 3GPP Work Item Description, Tech. Rep. R1-2212049 (2022-11), 2022.\textsuperscript{\ref{FN:ZT}}\textsuperscript{,}\label{FN:SS}}, and ToA-based\footnote{Intel. Corporation, “Evaluation of SL positioning,” 3GPP Work Item Description, Tech. Rep. R1-2212739 (2022-11), 2022.\label{FN:IT}} cannot meet the 0.5-meter accuracy with 90\% availability, even with 100 MHz bandwidth\cite{38_859}. Increasing to 400 MHz improves results across various approaches, yet time difference of arrival-based\textsuperscript{\ref{FN:IT}} techniques demand precise synchronization, while combined round-trip time and angle-of-arrival-based\textsuperscript{\ref{FN:ZT},\ref{FN:SS}} methods need accurate anchor locations, ample antenna elements, and line-of-sight paths to succeed at 0.5 meters.

To highlight the primary elements affecting SL positioning performance, we conducted simulations without BSs or RSUs. The setup and parameters appear in Fig. \ref{A_AS_90_ToA_AoA}.
\begin{itemize}
\item
As shown in the bottom left of Fig. \ref{A_AS_90_ToA_AoA}, \textbf{available spectrum resources play a major role in positioning accuracy} for ToA and AoA methods. With ideal synchronization between the target device and reference device, the chances of hitting 1-meter accuracy stand at 32\%, 75\%, and 95\% for 20 MHz, 40 MHz, and 100 MHz bandwidths, respectively. Since Release 18 requires 90\% service availability for positioning, we assessed the accuracy at the 90th percentile in Fig. \ref{A_AS_90_ToA_AoA}. In particular, time-of-arrival needs at least 100 MHz for 0.5-meter accuracy, while angle-of-arrival requires 40 MHz for 1.5-meter accuracy.
\item
The bottom left of Fig. \ref{A_AS_90_ToA_AoA} also demonstrates how \textbf{lack of synchronization between the target device and anchor sharply reduces positioning accuracy}. For instance, errors reach 6.2 meters, 2.5 meters, and 0.8 meters with 20 MHz, 40 MHz, and 100 MHz bandwidths. This means 1.5-meter errors are feasible with 100 MHz, but 0.5-meter errors remain out of reach even at that level. By comparison, the top right of Fig. \ref{A_AS_90_ToA_AoA} indicates that clock drift has a lesser impact on round-trip time plus angle-of-arrival errors, showing greater resilience to such imperfections than to synchronization issues.
\item
The \textbf{number of anchors greatly influences SL positioning accuracy} when relying on time-of-arrival measurements. The bottom right of Fig. \ref{A_AS_90_ToA_AoA} reveals that more anchors boost precision, with the improvement most evident at wider bandwidths. For example, doubling the number of anchors from 3 to 6 cuts errors by 92\%, 77\%, and 13\% at 100 MHz, 40 MHz, and 20 MHz, respectively. Such results underscore how broader bandwidths amplify the advantages of extra anchors in SL systems.
\end{itemize}

\subsection{Key finding 2: Performance disagreement between vendors} 
Vendors have yet to agree on the effectiveness of SL positioning. For example, one report\textsuperscript{\ref{FN:HW}} concludes that in highway settings for vehicle-to-everything communications, SL-only methods with 100 MHz bandwidth fall short of delivering 1.5-meter relative accuracy at 90\% availability. In contrast, another report\textsuperscript{\ref{FN:ZT}} asserts that this target is attainable under identical conditions. Comparable inconsistencies appear in urban grid environments, where evaluations differ on whether 100 MHz SL-only positioning can achieve the 1.5-meter threshold\textsuperscript{\ref{FN:HW},\ref{FN:ZT}}. Adding to the complexity, these evaluations often ignore elements like clock drift and timing mismatches, which hinders a full understanding of SL capabilities.

Such variations stem from differing simulation approaches used by vendors. While they typically share similar bandwidth allocations and techniques, including advanced resolution algorithms\textsuperscript{\ref{FN:ZT}}, differences in other setup details produce divergent outcomes. These include:
\begin{inparaenum}
\item \textbf{Lack of consensus on standard positioning approaches}. Given the varying effectiveness and demands of different techniques, as outlined in Table \ref{pos_methods}, selecting one or more as benchmarks is essential.
\item \textbf{Inconsistent configurations for antenna setups}. The quantity of antennas affects angle precision and enhances signal focusing, which in turn boosts overall positioning reliability. Yet, this factor is frequently neglected in assessments\textsuperscript{\ref{FN:ZT}}.
\item \textbf{Unaligned propagation models}. Vendors might apply, for instance, an adapted rural macro model\textsuperscript{\ref{FN:ZT}} based on TR 38.901 or an urban micro model\textsuperscript{\ref{FN:HW}} from TR 36.843. To resolve such inconsistencies, suggestions include customizing a standard propagation model to fit particular applications\textsuperscript{\ref{FN:ZT}}.
\end{inparaenum}

\section{Research Opportunities of SL Positioning}

\subsection{Multi-Hop Positioning}

The forthcoming 3GPP Release-19 standard aims to enhance coverage for SL communication and positioning by introducing multi-hop relay functionality \cite{chen20235g}. This innovation not only widens the service area but also enhances energy efficiency and the quality of service (QoS) \cite{chen20235g}. Multi-hop SL positioning is particularly beneficial in environments lacking a direct LoS path between the anchor and UE, where range and angle estimation tend to decline under NLoS conditions. By utilizing relay anchors, a LoS connection can be established, thereby improving positioning accuracy. Furthermore, multi-hop methods can enhance energy efficiency since lower transmission power is required for PRS to maintain equivalent signal quality and signal-to-noise ratio (SNR) in LoS situations compared to NLoS ones. However, the integration of multi-hop SL positioning faces problems of the relay selection and discovery, user authorization, and maintaining consistent QoS. As the number of hops increases, these challenges become more pronounced.

\subsection{AI/ML-Driven SL Positioning}
Recognizing the significant potential of artificial intelligence (AI) and machine learning (ML), the 3GPP has initiated standardization efforts (notably through TR 38.843\cite{38_843}), and categorized positioning approaches into two types: direct positioning and indirect positioning based on parameter estimation. Direct positioning predicts location coordinates directly from channel measurements\cite{38_215}, including SL PRS received power, SL PRS received signal strength indicator, etc. Traditional direct positioning is achieved by mapping channel measurements to user positions using pre-collected signal fingerprints, which is ineffective for SL positioning because the environment experienced by UE can vary rapidly, disrupting the direct relationship between channel measurements and position. Such challenge can be resolved by developing advanced AI/ML frameworks to learn the hidden pattern between channel measurements and positions\cite{xu2025enhanced}. In contrast, indirect positioning leverages AI/ML models to improve parameter estimation, which can learn complex, hidden features within channel measurements related to range and angle, leading to enhanced accuracy in localization. Despite these advantages, a major challenge remains in model design, especially within real-world wireless networks. Due to the limited computational power of UE, the AI/ML model should strike a balance between positioning performance and computational complexity, which remains an open problem\cite{pan2025ai}.

\subsection{Practical Issues of Sidelink Positioning}

Despite the promising capabilities of SL positioning, several practical issues need resolution before it can be effectively deployed in real-world systems. 

\begin{inparaenum}[\itshape a\upshape)]

\item Synchronization errors significantly affect time-based positioning methods, posing a major challenge, particularly in large-scale networks\cite{xu2025enhanced}. Among the advanced synchronization techniques, the Precision Time Protocol (PTP) shows potential for achieving nanosecond-level synchronization in SL positioning by providing high-precision time synchronization in packet-based networks. However, this requires additional hardware. Similarly, Reference Broadcast Synchronization (RBS) is effective in synchronizing multiple node clocks using a common time reference. Combining PTP and RBS can help further mitigate synchronization errors, leading to improved accuracy.

\item The efficiency and scalability of SL positioning networks face challenges related to deployment costs, energy consumption, and signaling overhead. Reports, such as those in \cite{38_859}, indicate that positioning accuracy can be enhanced with roadside units (RSUs), albeit requiring initial hardware investment. To reduce energy consumption, duty cycling is an effective strategy, allowing nodes, including RSUs, to switch to low-power modes when inactive, waking only for scheduled tasks. Additionally, energy harvesting presents a sustainable way to power operations by capturing radio frequency energy, potentially reducing or even eliminating battery dependence and supporting continuous operation. To address signaling overhead, advancements inspired by CSI feedback can be applied. For instance, compressing data for PRS allows signal reconstruction at the receiver, reducing transmission loads without compromising positioning accuracy.

\end{inparaenum}

\subsection{Integration with emerging technologies}
Fluid Antenna Systems (FAS) offer significant potential to enhance SL positioning in 6G networks, particularly for V2X applications, by dynamically reconfiguring antenna positions to maximize line-of-sight (LoS) probability, thereby improving the accuracy of positioning methods such as AoA, TDOA, and RTT. However, a key challenge in SL environments is the timely acquisition of accurate channel measurements while minimizing the computational complexity and signaling overhead, especially in NLOS and high-mobility scenarios (e.g., vehicles at 250 km/h per 3GPP Rel-19). Channel extrapolation, which predicts CSI across dynamic antenna configurations\cite{SSnet2025gao}, emerges as a highly promising technique to address these issues, enabling low-latency, high-accuracy positioning with reduced measurement overhead, thus aligning with the stringent requirements of 3GPP SL positioning standards.
%\cite{jiang2025towards,SSnet2025gao,jin2025linformer,gao2025enabling}

Reconfigurable intelligent surfaces (RIS) enhance SL positioning by manipulating radio wave propagation, thereby improving signal quality in NLOS settings. This allows for more precise ToA and AoA measurements, especially in dense urban environments. The passive nature of RIS reduces the energy consumption of SL devices, making it suitable for energy-efficient V2X applications. However, deploying RIS necessitates precise configuration and detailed CSI, which increases system complexity and overhead. Additionally, the cost and scalability of RIS panels present challenges for widespread adoption in SL networks.

\section{Conclusions}
In this article, we consolidated the technical aspects of SL positioning, which are fragmented across various 3GPP technical reports, and provided an overview of the latest standardization and features of SL positioning. We specifically detailed the network architecture, performance requirements, and physical layer design. We explored the controversial capability of SL positioning reported by different vendors in-depth, analyzing the underlying simulation settings. Our study demonstrated the significant impact of spectrum bandwidth and various imperfections on positioning performance using different methods, offering valuable simulation insights. Finally, we highlighted future research directions for SL positioning in line with the advancements in Rel-19 standardization of SL.
%\subsection{Acknowledgments}
%The authors would like to thank Dr Wanshi Chen of Qualcomm Inc., San Diego, United States, for insightful discussions on topics related to this work.

%简要介绍了UAV和RIS的论文，但也没有详细分析其协同效应，我觉得这并不是我们论文的主要内容

\bibliographystyle{IEEEtran}
\bibliography{Myreference}

\end{document}